\documentclass[aps,pre,reprint,superscriptaddress,noofootinbib,showpacs,floatfix]{revtex4-1}

\usepackage{graphicx}
\usepackage{latexsym}
\usepackage{amsmath}
\usepackage{multirow}

\usepackage[usenames]{color}

\newcommand{\lb}{l_\text b}
\renewcommand{\vec}[1]{\mathbf{#1}}
\newcommand{\uvec}[1]{\vec{\widehat{#1}}}

\begin{document}

\title{Twist-bend coupling, twist waves and the shape of DNA loops}

\author{S. K. Nomidis}

\affiliation{Laboratory for Soft Matter and Biophysics, KU Leuven,  
Celestijnenlaan 200D, 3001 Leuven, Belgium}

\affiliation{Flemish Institute for Technological Research (VITO), Boeretang 
200, B-2400 Mol, Belgium}

\author{M. Caraglio}

\affiliation{Laboratory for Soft Matter and Biophysics, KU Leuven,  
Celestijnenlaan 200D, 3001 Leuven, Belgium}

\affiliation{Institut f\"{u}r Theoretische Physik, Universit\"{a}t
Innsbruck, Technikerstra{\ss}e  21A, A-6020 Innsbruck, Austria}

\author{M. Laleman}

\affiliation{Laboratory for Soft Matter and Biophysics, KU Leuven,  
Celestijnenlaan 200D, 3001 Leuven, Belgium}

\author{K. Phillips}

\affiliation{Laboratory for Soft Matter and Biophysics, KU Leuven,  
Celestijnenlaan 200D, 3001 Leuven, Belgium}

\author{E. Skoruppa}

\affiliation{Laboratory for Soft Matter and Biophysics, KU Leuven,  
Celestijnenlaan 200D, 3001 Leuven, Belgium}

\author{E. Carlon}

\affiliation{Laboratory for Soft Matter and Biophysics, KU Leuven,  
Celestijnenlaan 200D, 3001 Leuven, Belgium}

\begin{abstract}
By combining analytical and numerical calculations, we investigate the
minimal-energy shape of short DNA loops of approximately $100$ base pairs
(bp).  We show that in these loops the excess twist density oscillates
as a response to an imposed bending stress, as recently found in DNA
minicircles and observed in nucleosomal DNA. These twist oscillations,
here referred to as twist waves, are due to the coupling between twist
and bending deformations, which in turn originates from the asymmetry
between DNA major and minor grooves. We introduce a simple analytical
variational shape, that reproduces the exact loop energy up to the
fourth significant digit, and is in very good agreement with shapes
obtained from coarse-grained simulations.  We, finally, analyze the loop
dynamics at room temperature, and show that the twist waves are robust
against thermal fluctuations. They perform a normal diffusive motion,
whose origin is briefly discussed. 
\end{abstract}

\date{\today}

\maketitle

\section{Introduction}

DNA often forms loops due to the action of proteins, which bind at
two distant sites along its sequence and bring them in close contact
with each other~\cite{albe02}. DNA loops play an important role in
many biological processes such as transcription, recombination and
duplication. The loop length can range from 100 base pairs (bp), in the
{\sl lac} and {\sl gal} operons of E. Coli \cite{cour13}, to tenths
of thousands bp in more complex organisms \cite{kriv12}. For lengths
much longer than the DNA persistence length, $\lb \approx 50$~nm ($150$
bp), entropic contributions dominate, leading to strong fluctuations
in the shape, while in the opposite limit, namely on loops of length
comparable to or smaller than $\lb$, the loop assumes approximately its
minimal-energy shape.  Several studies have focused on the latter regime,
not only due to its biological relevance, but also because it allows to
investigate the mechanics of highly-deformed DNA \cite{bala99, kuli03,
sank05, zhan06, lee10, wils10, cher11, vafa12, le14, chen14, mull15}.

Mechanical models used for short loops typically neglect
sequence-dependent effects, and describe DNA as an isotropic
continuous elastic rod. This description fails to account for a
coupling interaction present in real DNA, that connects the bending
and twisting degrees of freedom, and originates from the asymmetry of
the DNA grooves~\cite{mark94}. The effect of such a twist-bend coupling
interaction on the behavior of DNA at various scales has been recently
discussed \cite{nomi17, skor17, skor18, nomi19, cara19}. The aim of this
paper is to investigate its influence on the structure of short DNA loops.
Starting from a simple wormlike chain (WLC) model, in which only the
bending degrees of freedom are taken into account, we construct a simple
variational ansatz for the loop shape, which we refer to as harmonic
loop. This ansatz reproduces very accurately the exact loop shape, which
is expressed in terms of elliptic integrals \cite{yama72}, while the
exact loop energy is reproduced by the harmonic loop ansatz up to four
significant digits. The advantage of the variational solution is that it
involves simple trigonometric functions, from which various properties of
the loop can be easily obtained.  Combining with the results of previous
work \cite{skor18}, we have extended the variational solution to more
complex DNA models, with anisotropic bending and twist-bend coupling. The
comparison with numerical simulations of various coarse-grained model of
DNA shows that the harmonic-loop approximation performs extremely well
in all cases. As such, it provides a basis for further analysis of the
equilibrium and kinetic properties of the DNA loops, which are briefly
addressed at the end of this paper.

\section{DNA elasticity and twist-bend coupling}
\label{sec:elastic}

The simplest continuum model of DNA is the WLC, in which a configuration
is described by the tangent vector $\uvec{e}_3(s)$, where $s$ is the
curvilinear contour coordinate ($0 \leq s \leq L$) and $L$ the total
length. The energy takes the following form:
\begin{equation}
\label{eq:WLC}
\beta E = \frac{\lb}{2} \int_0^L \text ds 
\left(\frac{\text d \uvec{e}_3}{\text d s} \right)^2,
\end{equation} 
where $\beta = 1/k_\text BT$ is the inverse temperature and $\lb$ the
bending persistence length. To include twist, the model can be extended
by introducing two additional unit vectors $\uvec{e}_1$ and $\uvec{e}_2$,
such that $\{\uvec e_1, \uvec e_2, \uvec e_3\}$ forms an orthonormal
basis. The vectors $\uvec{e}_1$ and $\uvec{e}_2$ keep track of the
relative rotation around the $\uvec{e}_3$ axis of neighboring points. In
DNA, by convention, $\uvec{e}_2$ connects the backbones of the two strands
and $\uvec{e}_1$ points towards the major groove. A generic configuration
of the twisted rod can be then parametrized in terms of infinitesimal
rotations connecting the orthonormal frame $\{\uvec{e}_1, \uvec{e}_2,
\uvec{e}_3\}$ in $s$ to a neighboring frame in $s+\text ds$. This rotation
can be mathematically cast into the following differential equation
\begin{equation}
\label{eq:rot_DE}
\frac{\text{d} \uvec{e}_i}{\text{d} s} = \left(\vec{\Omega}+
\omega_0 \uvec{e}_3\right) \times \uvec{e}_i,
\end{equation}
where $i=1,2,3$, and $\omega_0\approx 1.75$ rad/nm is the intrinsic
helical twist density. The solution of the previous equation for
$\vec\Omega = \vec 0$ corresponds to a twisted straight rod ($\uvec{e}_3
= \text{const}.$), in which $\uvec{e}_1$ and $\uvec{e}_2$ rotate with
angular frequency $\omega_0$. One defines the three components of the
vector $\vec{\Omega}$ along the frame as $\Omega_i \equiv \vec{\Omega}
\cdot \uvec{e}_i$. Here, $\Omega_1$ and $\Omega_2$ denote the bending
densities along the two main axes of the molecule and $\Omega_3$ the
excess twist density. From the analysis of the symmetry of a DNA molecule,
Marko and Siggia derived the following continuum model~\cite{mark94}
\begin{equation} 
\beta E = \frac{1}{2}
\int_0^L \! \text{d} s \left( A_1 \Omega_1^2 + A_2 \Omega_2^2 + C \Omega_3^2 + 
 2G  \Omega_2 \Omega_3 \right).
\label{eq:model}
\end{equation}
The parameters $A_1$ and $A_2$ are the stiffnesses associated with
bending over the backbone and the grooves, respectively, while $C$ is the
intrinsic twist stiffness. Finally, the twist-bend coupling term $G$ leads
to a correlation of the strain fields $\Omega_2$ and $\Omega_3$. 
Note that from Eq.~\eqref{eq:rot_DE} one obtains
\begin{equation}
\kappa^2 \equiv \left(\frac{\text{d} \uvec{e}_3}{\text{d} s} \right)^2 
= \Omega_1^2 + \Omega_2^2,
\label{eq:kappa}
\end{equation}
where $\kappa$ is the curvature. Using this relation, and setting $A_1
= A_2 = \lb$, while neglecting twist degrees of freedoms, one directly
sees that model~\eqref{eq:model} reduces to \eqref{eq:WLC}.

The effect of twist-bend coupling on the mechanical properties of DNA has
been investigated in a few recent papers \cite{nomi17, skor17, skor18,
cara19, nomi19}. An interesting consequence of $G \neq 0$ is the existence
of twist oscillations in curved DNA \cite{skor18}. In Ref.~\cite{skor18}
the following minimal-energy shape of a minicircle was derived
\begin{equation}\label{om_tr}
\begin{aligned}
\Omega_1 &= \frac{\lb}{A_1}\frac{\sin (\omega_0 s)}{R},\\
\Omega_2 &= \frac{\lb}{\widetilde{A}_2}\frac{\cos (\omega_0 s)}{R},\\
\Omega_3 &= -\frac{G}{C} \Omega_2,
\end{aligned}
\end{equation}
where $R$ is the average circle radius and
\begin{equation}
\widetilde{A}_2 = A_2 \left( 1 - \frac{G^2}{A_2 C}\right),
\label{shiftA2}
\end{equation}
an effective bending stiffness. Finally, $\lb$ is the bending persistence
length, which within model~\eqref{eq:model} is given by
\begin{equation}
\frac{1}{\lb} = \frac{1}{2} \left(
\frac{1}{A_1}+\frac{1}{\widetilde{A}_2} \right), 
\label{eq:def_lb}
\end{equation}
i.e., the harmonic mean of the two stiffnesses~\cite{nomi17}. The
oscillations in the bending densities $\Omega_1$ and $\Omega_2$
arise from the geometrical constraints of the system, which in
turn induce oscillations in the twist density $\Omega_3$ from the
presence of twist-bend coupling ($G\neq0$). These twist waves have
been indeed observed in X-ray crystallographic structures of DNA
bound to histone proteins \cite{skor18}. While Eq.~\eqref{om_tr}
describes a torsionally-relaxed minicircle, this solution has also
been recently extended to minicircles which are either over- or
undertwisted~\cite{cara19}.

\section{Harmonic loops in the WLC model}
\label{sec:HL}

Considerable attention has been devoted to the study of structural and
dynamical properties of DNA loops in the past two decades \cite{bala99,
kuli03, sank05, zhan06, lee10, wils10, cher11, vafa12, le14, chen14,
mull15}. Already in the early 70s, Yamakawa and Stockmayer \cite{yama72}
discussed the minimal-energy configuration of a semiflexible loop within
the framework of the isotropic WLC [Eq.~\eqref{eq:WLC}]. The exact shape
thus obtained is expressed in terms of inverse elliptic integrals, and
its derivation is outlined in Appendix \ref{appA}. Although exact, these
expressions are not easy to handle, therefore simpler approximate loop
configurations have also been considered. For instance, in the context
of DNA looping in the nucleosome, Kulic and Schiessel \cite{kuli03}
introduced a so-called circle-line approximation, in which the DNA
conformation is built from straight segments and arcs of circles. A
similar approach was followed by Sankararaman and Marko \cite{sank05},
who studied DNA loops under tension.  In the same spirit, we introduce
here a different approximate shape for the loop, which we refer to as
harmonic loop.  Though still simple, it is found to be more accurate, and
will allow us to construct a full analytical shape for the anisotropic
model \eqref{eq:model}.  From that, we can directly estimate several
quantities of interest, such as twist oscillations, curvature variation
and minimal energy.

As the problem is two-dimensional, one can describe the shape of the
loop using a single parameter $\theta(s)$, defined as the angle
the tangent $\uvec e_3$ forms with the x-axis:
\begin{equation}
\uvec{e}_3 = \cos \theta \, \uvec{x} + \sin \theta \, \uvec{y},
\label{eq:e3_teardrop}
\end{equation}
where the unit vectors $\uvec{x}$ and $\uvec{y}$ lie on the plane of
the loop. Since no rotational constraints are applied at the
loop endpoints, the curvature $\kappa = |\text d\theta/\text d s|$
[see Eq.~\eqref{eq:kappa}] must vanish at the boundaries, hence
\begin{equation}
\frac{\text{d} \theta (0)}{\text{d}s} =
\frac{\text{d} \theta (L)}{\text{d}s} = 0.
\label{eq:bc}
\end{equation}
A simple ansatz fulfilling these boundary conditions is
\begin{equation}
\frac{\text{d} \theta^{(1)}}{\text{d} s} = \frac{\pi c_1}{L} 
\sin \left( \frac{\pi s}{L} \right),
\label{eq:HL}
\end{equation}
which we refer to as first-order harmonic loop. Here, the dimensionless
constant $c_1$ can be fixed by requiring that the endpoints coincide.
This constraint is discussed in Appendix~\ref{appB} and can be cast in the
form $J_0(c_1)=0$, with $J_0$ being the zeroth-order Bessel function of
the first kind [see Eq.~\eqref{eq:fix_c1}]. The parameter $c_1$ is, thus,
the first root of $J_0$, which is known to a high degree of accuracy,
$c_1 = 2.40482556$.

\begin{figure}[t]
\centering\includegraphics{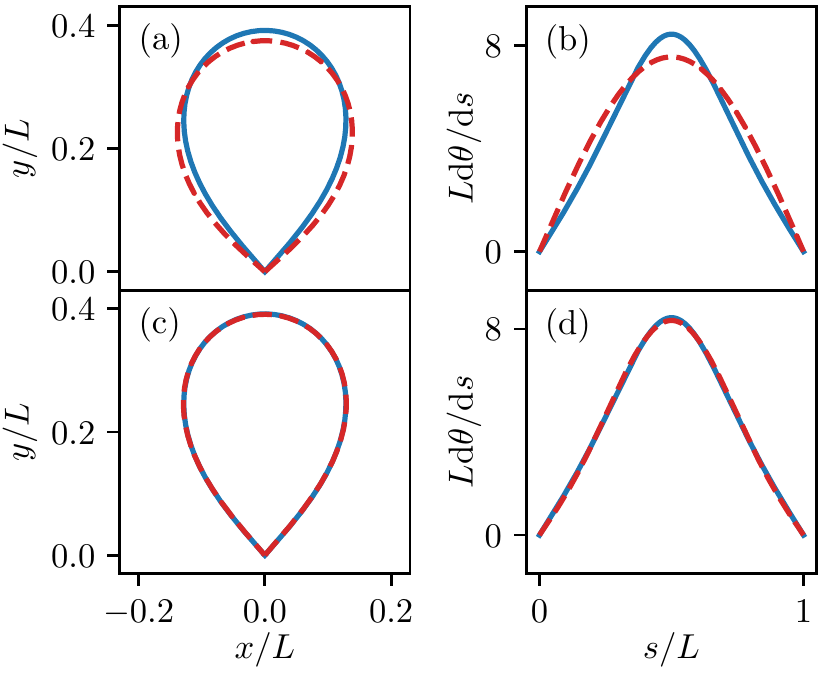}
\caption{Minimal-energy loop shape (left) and curvature as a function of
the rescaled arc-length coordinate $s/L$ (right). Blue solid lines are
obtained from the exact solution~\eqref{eq:FF} \cite{yama72}, while red
dashed lines show (a,b) the first- and (c,d) third-harmonic solutions,
corresponding to Eqs.~\eqref{eq:HL} and \eqref{eq:HL3}, respectively.}
\label{fig:1st_harmonic}
\end{figure}

Figure~\ref{fig:1st_harmonic}(a) shows with a red dashed line the
loop shape, obtained by integrating Eq.~\eqref{eq:HL}, plugging into
Eq.~\eqref{eq:e3_teardrop} and integrating once more. The red dashed line
in Fig.~\ref{fig:1st_harmonic}(b) shows a plot of Eq.~\eqref{eq:HL},
multiplied by the loop length $L$, so as to render it dimensionless.
For comparison, the same graphs show as blue solid lines the shapes
and curvature corresponding to the exact solution~\eqref{eq:FF}
\cite{yama72}. There is a reasonable overall agreement between the exact
solution and first-harmonic approximation. Plugging Eqs.~\eqref{eq:HL}
and \eqref{eq:e3_teardrop} into Eq.~\eqref{eq:WLC}, we find the following
energy for the harmonic loop
\begin{equation}
\beta E^{(1)}_\text{HL} = \left(\frac{\pi c_1}{2} \right)^2 \frac{\lb}{L}
= 14.2694 \, \frac{\lb}{L},
\label{eq:enHL1}
\end{equation}
which is very close to the exact value $\beta E_\text{exact} = 14.0550 \,
(\lb/L)$ of the Yamakawa-Stockmayer solution \cite{yama72}.  The value
in Eq.~\eqref{eq:enHL1} improves upon the ``circle-line'' approximation
of Ref.~\cite{kuli03}, which gives $\beta E_\text{CL} = 15.70 \, (\lb/L)$.

Our approximation scheme can be systematically improved by including 
higher-order terms. We consider the following variational ansatz
\begin{equation}
\frac{\text{d} \theta^{(3)}}{\text{d} s} = \frac{\pi}{L}
\left[  
c_1 \sin \left( \frac{\pi s}{L}  \right) + 
c_3 \sin \left( \frac{3\pi s}{L} \right)
\right],
\label{eq:HL3}
\end{equation}
which extends Eq.~\eqref{eq:HL} by adding an additional harmonic,
consistent with the symmetry of the solution (we require $\text
d\theta/\text ds$ to be symmetric around $s=L/2$, which excludes all even
harmonics with frequency $2\pi n/L$ and $n$ integer). The two parameters
$c_1$ and $c_3$ are now fixed by requiring both the closure of the loop
and the energy minimization of Eq.~\eqref{eq:WLC}. We, thus, find the
values $c_1=2.3703$ and $c_3=-0.2808$ (more details can be found in
Appendix~\ref{appB}). The resulting shape and rescaled curvatures are
shown as dashed red lines in Fig.~\ref{fig:1st_harmonic}(c) and (d),
revealing an excellent agreement with the exact solution (solid blue
line). The energy is found to be
\begin{equation}
\beta E^{(3)}_\text{HL} =  \frac{\pi^2 \left(c_1^2 + c_3^2\right)}{4} 
\, \frac{\lb}{L} = 14.0572 \, \frac{\lb}{L},
\label{eq:enHL}
\end{equation}
which matches the exact solution up to four significant digits.
The ansatzes of Eqs.~\eqref{eq:HL} and \eqref{eq:HL3} can be extended
to the case where the end-points are fixed at some finite distance~$d$
(see Appendix~\ref{appB}). Table \ref{table} summarizes the optimal values
of the coefficients $c_1$ and $c_3$ for the first- and third-harmonic
approximations for some selected values of $d$. A comparison between
the exact results and the harmonic approximations shows that the latter
become even better with increasing $d$, i.e., as the distance between
the end-points increases (for more details, see Appendix~\ref{appB}).

\begin{table}[t]
\caption{Comparison between harmonic-loop energies and the exact
values for three different ``loop'' geometries with varying distance
$d$ between the endpoints. $d=0$ corresponds to the closed loops of
Fig.~\ref{fig:1st_harmonic}.}
\begin{tabular}{c|c@{\hspace*{0.4cm}}c@{\hspace*{0.4cm}}c@{\hspace*{0.4cm}}c
@{\hspace*{0.4cm}}c@{\hspace*{0.3cm}}}
\hline
 d & & $c_1$ & $c_3$ & $\beta E L/\lb$ &  Error(\%)\\
\hline
\multirow{3}{*}{$0$}
 &1st & 2.4048 & --  	 & 14.2694 & 1.53 \\
 &3rd & 2.3703 & -0.2808 & 14.0572 & 0.02 \\
 &Exact &      &         & 14.0550 &  \\
\hline
\multirow{3}{*}{$L/10$}
 &1st & 2.2187 & --  	 & 12.1459 & 0.98 \\
 &3rd & 2.1977 & -0.2135 & 12.0295 & 0.02 \\
 &Exact &      &         & 12.0286 &  \\
\hline
\multirow{3}{*}{$L/5$}
 &1st & 2.0415 & --  	& 10.2837 & 0.63 \\
 &3rd & 2.0288 & -0.1612 & 10.2198 & $<$0.01 \\
 &Exact & --   & --      & 10.2194 &  \\
\hline
\end{tabular}
\label{table}
\end{table}

From the simple analytical, but accurate, form of the loop shape one can
easily obtain several estimates of the loop properties. Let us consider,
for instance, the curvature $\kappa(s) = \text d\theta/ \text d s$, which
reaches its maximum value $\kappa_\text{max}$ at the apex $s=L/2$ of
the loop. From the third-harmonic approximation~\eqref{eq:HL3}, one finds
\begin{equation}\label{eq:kmax}
\kappa_{\max} = \frac{\pi}{L} \left(c_1 - c_3\right).
\end{equation}
Next, we wish to estimate the curvature~$\kappa_\text{turn}$ at a distance
of one helical repeat from the apex, i.e., at $s = L/2 + 2\pi/\omega_0$.
Again, from Eq.~\eqref{eq:HL3} this is found to be
\begin{equation}\label{eq:kturn}
 \kappa_\text{turn} = \frac{\pi}{L} \left( c_1\cos\alpha - 
c_3\cos3\alpha \right),
\end{equation}
where we introduced the angle $\alpha \equiv 2\pi^2/\omega_0L$. Finally,
combining Eqs.~\eqref{eq:kmax} and \eqref{eq:kturn}, one finds the
maximal curvature drop
\begin{equation}
 \frac{\Delta\kappa}{\kappa_\text{max}} \equiv 
 \frac{\kappa_\text{max} - \kappa_\text{turn}}{\kappa_\text{max}}
 = 1 - \frac{c_1\cos\alpha - c_3\cos3\alpha}{c_1-c_3}.
\label{eq:drop_kappa}
\end{equation}
Consulting Table~\ref{table} for $d = 0$ and considering a DNA loop of
100~bp ($L=34$~nm), yields $\kappa_\text{max} = 0.24\, \text{nm}^{-1}$
and $\Delta\kappa/\kappa_\text{max} \approx 0.097$.

\section{Minimal-energy configuration of DNA loops: Modulated twist waves}
\label{sec:T0}

\subsection{Analytical results}

The analysis of Eq.~\eqref{eq:drop_kappa} for a loop of $100$~bp reveals
a rather modest curvature drop at the scale of the helical-repeat
length. This allows us to use Eq.~\eqref{om_tr}, derived for a minicircle
of average radius $R$, by replacing $1/R$ with the modulated harmonic-loop
curvature. For instance, combining Eq.~\eqref{eq:HL} with \eqref{om_tr}
yields the first-harmonic solution
\begin{equation}
\begin{aligned}
\Omega^{(1)}_1 &= 
\frac{\lb}{A_1} \, \frac{\pi c_1}{L} 
\sin \left( \frac{\pi s}{L} \right) \sin (\omega_0 s+ \phi ), \\
\Omega^{(1)}_2 &= 
\frac{\lb}{\widetilde{A}_2} \frac{\pi c_1}{L}
\sin \left( \frac{\pi s}{L} \right) \cos (\omega_0 s+ \phi ), \\
\Omega^{(1)}_3 &= -\frac{G}{C} \Omega^{(1)}_2,
\end{aligned}
\label{om_modulated}
\end{equation}
where a phase constant $\phi$ has been added, accounting for the torsional
freedom of DNA at the boundaries (torsionally-unconstrained ends).
Similarly, one can combine Eq.~\eqref{om_tr} with Eq.~\eqref{eq:HL3},
so as to construct a more accurate approximation
\begin{equation}
\begin{aligned}
\Omega^{(3)}_1 &= \frac{\lb}{A_1} \frac{\pi}{L} 
\left[ c_1 \sin \left( \frac{\pi 
s}{L} \right) + c_3 \sin\left(\frac{3\pi s}{L}\right)\right] \sin 
(\omega_0 s+ \phi ), \\
\Omega^{(3)}_2 &= \frac{\lb}{\widetilde{A}_2} \frac{\pi}{L} \left[ c_1 \sin 
\left( \frac{\pi s}{L} \right) + c_3 \sin\left(\frac{3\pi 
s}{L}\right)\right] \cos (\omega_0 s+ \phi ), \\
\Omega^{(3)}_3 &= - \frac{G}{C} \Omega^{(3)}_2.
\end{aligned}
\label{om_modulated_3}
\end{equation}
Similar to the minicircle case [Eqs.~\eqref{om_tr}], one notices the
emergence of twist waves, originating from twist-bend coupling ($G \neq
0$). In this case, however, these are modulated by the varying curvature,
which vanishes at the loop edges and is maximal at the loop apex.

By plugging Eqs.~\eqref{om_modulated} into Eq.~\eqref{eq:model} and
performing the integration in $s$, one can compute the total energy of
the loop (see Appendix~\ref{appC} for details)
\begin{equation}
\beta E^{(1)}_\text{HL} = \left(\frac{\pi c_1}{2} \right)^2 \frac{\lb}{L}
+ \beta \Delta E(\phi).
\label{eq:Ephi}
\end{equation}
This expression is identical to Eq.~\eqref{eq:enHL1} with the addition
of a boundary term $\Delta E(\phi)$ depending on the phase $\phi$. One
can show that this term is negligible for loops of about 100~bp, such as
those considered here ($|\Delta E|/E^{(1)}_\text{HL} \sim 10^{-4}$). We,
thus, conclude that the energy is quasi-degenerate, corresponding to an
invariance of the double helix with respect to a global rotation~$\phi$
around its axis. A similar conclusion holds for the third-harmonic
approximation.

\begin{figure}[t]
\centering\includegraphics{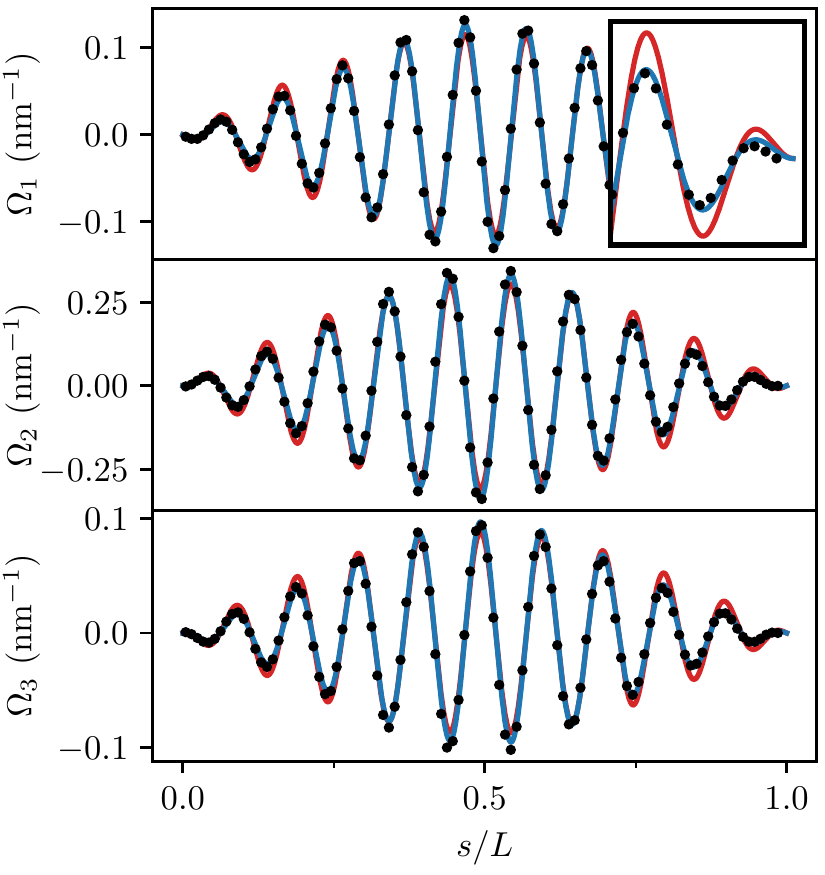}
\caption{Plots of the bending ($\Omega_1$ and $\Omega_2$) and twist
($\Omega_3$) densities of a closed triad-model loop, as functions of
the reduced arc-length $s/L$. The black circles are low-temperature
Monte Carlo simulations of the triad model (with $104$~bp), while the
red and blue solid lines are the first- [Eq.~\eqref{om_modulated}] and
third-harmonic [Eq.~\eqref{om_modulated_3}] ansatzes, respectively. The
inset zooms into the one end of the loop, and reveals that the third-harmonic 
ansatz more accurately reproduces the data. The stiffness
constants were chosen as $A_1=81~\text{nm}$, $A_2=39~\text{nm}$,
$C=105~\text{nm}$ and $G=30~\text{nm}$, calculated in Ref.~\cite{skor17}
from the oxDNA2 model.  The global phase $\phi=4.02$~rad is the only
fitting parameter.}
\label{fig:deformation_parameters}
\end{figure}

\begin{figure*}[t]
\centering\includegraphics{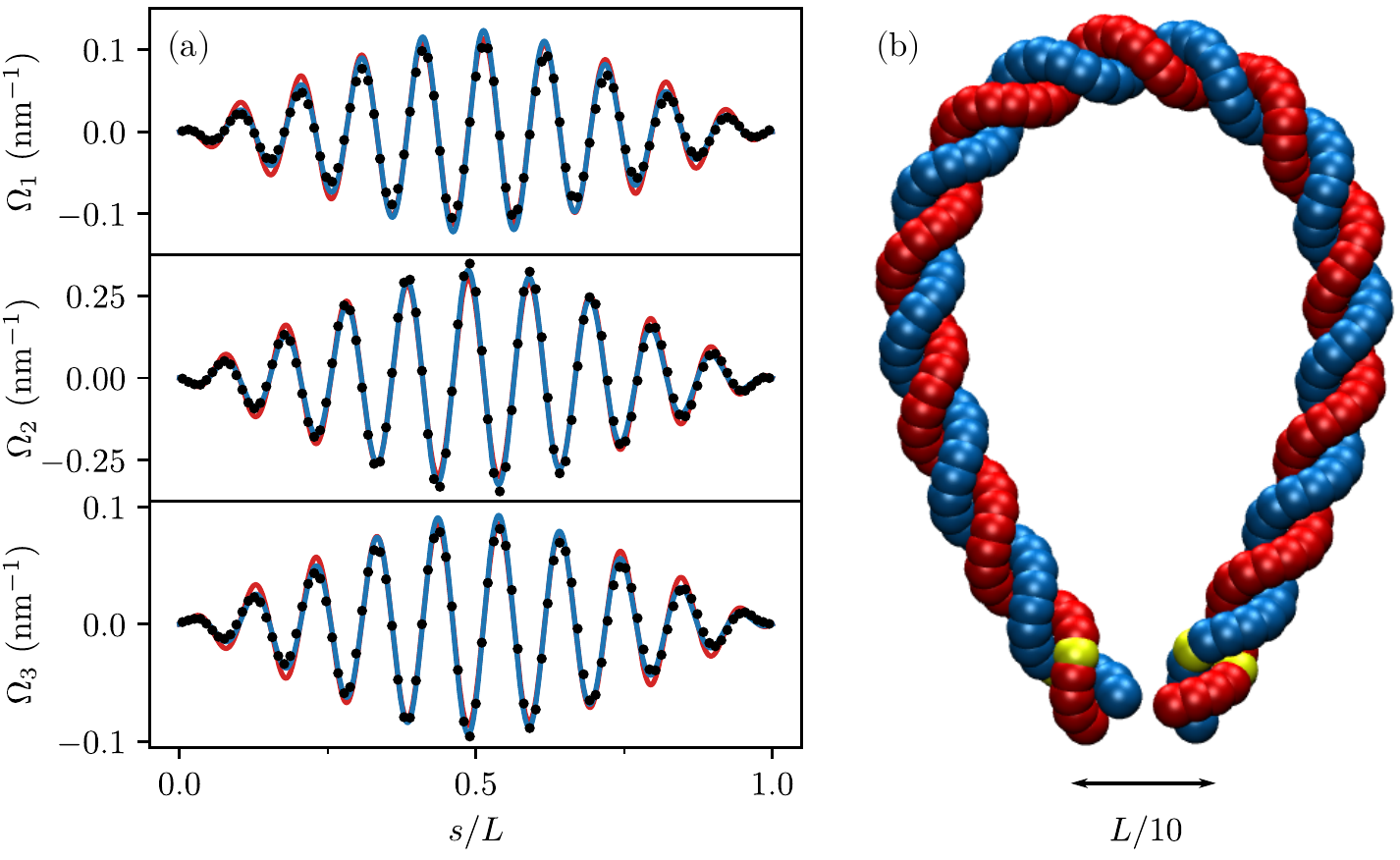}
\caption{(a) Similar to Fig.~\ref{fig:deformation_parameters}, but
with the data obtained from low-temperature oxDNA2 simulations (black
points). For the harmonic approximations [Eqs.~\eqref{om_modulated} and
\eqref{om_modulated_3}, shown with red and blue lines, respectively]
we used the same values as in Fig.~\ref{fig:deformation_parameters},
extracted from oxDNA2 simulations \cite{skor17}. Once more, the global
phase $\phi = 1.70$~rad is the only fitting parameter. (b) Actual
configuration of the loop from which the plots in (a) were produced.
The yellow points indicate the bp that were tethered to two fixed points
in space by means of harmonic springs, effectively fixing the end distance
(center-of-mass distance between the two pairs of yellow beads) at $d
= L/10$.}
\label{fig:oxdna_loop}
\end{figure*}

Finally, from Eq.~\eqref{om_modulated} [or Eq.~\eqref{om_modulated_3}]
one can estimate the curvature at the loop apex
\begin{equation}
\widetilde{\kappa}_{\max} \equiv 
\left. \sqrt{\Omega_1^2 + \Omega_2^2}\right|_{s=L/2},
\end{equation}
and compare it with the corresponding WLC loop curvature
$\kappa_{\max} = \pi c_1/L$ [Eq.~\eqref{eq:kmax}], derived in
Sec.~\ref{sec:HL}. $\widetilde{\kappa}_{\max}$ depends on $\phi$
and, using some simple algebra, one can show that the ratio
$\widetilde{\kappa}_{\max}/\kappa_{\max}$ is bounded within the interval
\begin{equation}
\frac{2 \widetilde{A}_2}{A_1+\widetilde{A}_2} 
\leq \frac{\widetilde{\kappa}_{\max}}{\kappa_{\max}} \leq
\frac{2 A_1}{A_1+\widetilde{A}_2},
\label{eq:kappaL2}
\end{equation}
(note that the same expression is valid both for the first- and
third-harmonic approximation).  Equation~\eqref{eq:kappaL2} shows that
$\widetilde{\kappa}_{\max}$ does not, in general, coincide with the
WLC loop apex curvature $\kappa_{\max}$. The latter has been obtained
for a perfectly planar loop, however the solution~\eqref{om_tr} [and,
thus, also \eqref{om_modulated} and \eqref{om_modulated_3}] describes an
almost-planar curve with small off-planar oscillations \cite{skor18},
which are induced by the combined effect of bending anisotropy and
twist-bend coupling.

\subsection{Numerical analysis}
 
In order to test the validity of Eqs.~\eqref{om_modulated} and
\eqref{om_modulated_3}, we performed Monte Carlo simulations of a
``triad model'', which is derived from a direct discretization of
Eq.~\eqref{eq:model}.  A DNA molecule with $N$ base pairs is represented
by $N$ beads, each carrying a set of three orthogonal unit vectors
forming the triad $\lbrace \uvec{\bf e}_1, \uvec{\bf e}_2, \uvec{\bf e}_3
\rbrace$, with $\uvec{\bf e}_3$ being the tangent, and hence pointing
towards the next bead. Consecutive beads are separated by a fixed distance
$a=0.34$~nm, corresponding to the average base pair distance of DNA.
The simulations are performed at sufficiently-low temperature, so that
the system converges to its lowest-energy state (more details can be
found in Ref.~\cite{cara19}). Figure~\ref{fig:deformation_parameters}
shows the bending densities ($\Omega_1$, $\Omega_2$) and
excess twist density ($\Omega_3$) as functions of the rescaled
arc-length coordinate $s/L$ obtained from Monte Carlo simulations
of a loop of $104$~bp (black circles). The lines are plots of
Eqs.~\eqref{om_modulated} (red) and Eqs.~\eqref{om_modulated_3}
(blue). The only adjustable parameter is the global phase $\phi$,
as the stiffness constants $A_1$, $A_2$, $C$ and $G$ are input
parameters (see caption of Fig.~\ref{fig:deformation_parameters}),
while $c_1$ and $c_3$ are the universal constants given in
Table~\ref{table}. Figure~\ref{fig:deformation_parameters} shows
an excellent agreement between both harmonic approximations
and the Monte Carlo data. As in the case of WLC loops shown in
Fig.~\ref{fig:1st_harmonic}, the first-harmonic approximation
overestimates the curvature at the loops ends (see inset of
Fig.~\ref{fig:deformation_parameters}). The maximal excess twist in a
loop with $104$~bp is $\max|\Omega_3| \approx 0.1~\text{nm}^{-1}$, which,
compared to the average intrinsic twist $\omega_0=1.75~\text{nm}^{-1}$,
corresponds to a deviation of 6\% from $\omega_0$. Finally, note
that the maximum curvature $\kappa_\text{max} = 0.24$~nm$^{-1}$,
predicted by Eq.~\eqref{eq:kmax}, is once more substantially lower
than the total curvature in the middle of the loop of the triad model,
which from the numerical data is found to be $\widetilde{\kappa}_{\max}
=0.34$~nm$^{-1}$. This is due to off-planar oscillations along the loop,
as discussed above [Eq.~\eqref{eq:kappaL2}].

To further corroborate the harmonic-loop approximations, we performed
low-temperature ($T = 1$~K) computer simulations of 100-bp oxDNA
loops (Fig.~\ref{fig:oxdna_loop}). OxDNA is a coarse-grained
model, which describes DNA as two intertwined strands of rigid
nucleotides~\cite{ould10}. For the simulations we used the latest
version oxDNA2~\cite{snod15}, which was recently found to have a
substantially-nonzero twist-bend coupling constant~\cite{skor17}.
The electrostatic interactions are implicitly modelled through
a Debye-H\"uckel potential. The initial configuration was a
torsionally-relaxed helix, with the molecular axis having the shape of
Eq.~\eqref{eq:HL3}. To avoid undesired interactions between the two DNA
ends, we constrained them at a nonzero distance $d = L/10$ by means of
strong harmonic bonds.  These bonds connect the center of mass of the
fifth base pair at each end (yellow beads in Fig.~\ref{fig:oxdna_loop}b)
with a fixed point in space (distance~$d$ between the two centers of
mass).  This is to prevent end-point denaturation effects, which can
affect in particular the room temperature simulations, such as those
presented in Section~\ref{sec:T300}. The simulations were performed with
the recently-developed LAMMPS~\cite{plim95} implementation of the oxDNA
model~\cite{henr18}. Figure~\ref{fig:oxdna_loop}a shows $\Omega_i$ as
functions of the arc-length parameter, and once more reveals an excellent
agreement with the harmonic approximations. Again, the only free parameter
is the global phase $\phi$, as the stiffness constants $A_1$, $A_2$, $C$
and $G$ have already been calculated for oxDNA2 in Ref.~\cite{skor17},
while $c_1$ and $c_3$ were taken from Table~\ref{table} with $d = L/10$.

\begin{figure}[t]
\centering\includegraphics{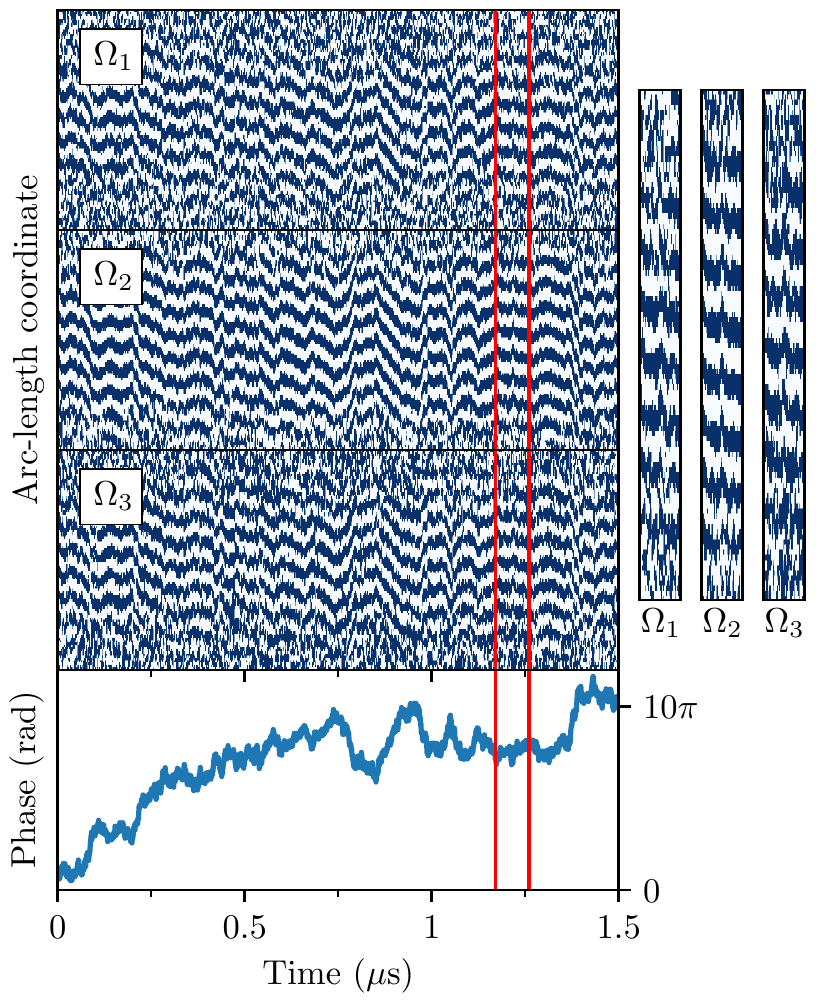}
\caption{Kymogram plots of the time evolution of the deformation
densities $\Omega_i$ along an oxDNA loop (top three panels), together
with the corresponding global phase (bottom panel), obtained from
room-temperature simulations. The right subplots show a zoom in the
region marked by red lines, revealing that the correlations predicted
by Eqs.~\eqref{om_modulated} and \eqref{om_modulated_3} survive the
effect of thermal fluctuations. In the kymographs, the blue and white
regions correspond to positive and negative values of the $\Omega_i$'s,
respectively.}
\label{fig:kin}
\end{figure}

\section{Effect of Thermal Fluctuations}
\label{sec:T300}

Having analyzed the minimal-energy conformation of DNA loops, we
now consider the effect of thermal fluctuations. For this purpose,
we simulated oxDNA loops using the same setup as discussed in Section
\ref{sec:T0}, with the only difference that the temperature now is $T =
300$~K. Figure~\ref{fig:kin} shows kymogram traces of $\Omega_i$ along
the DNA loop as functions of time. To better distinguish regions of
predominantly-positive $\Omega_i$ from those of negative $\Omega_i$,
we applied a Gaussian filter on both the spatial ($3$-bp variance) and
temporal direction (4-frame variance), and used a binary color code
(white for negative and blue for positive values). At the loop ends,
the image is more blurred due to the thermal fluctuations dominating
over the wave amplitude, as expected from the vanishing curvature.
In the central region, however, clear wave patterns are visible in the
bending and twist deformations, with a period following that of the DNA
double helix, in line with the ground-state solutions~\eqref{om_modulated}
and \eqref{om_modulated_3}. In order to better compare the relative phase
among $\Omega_i$, the side figure shows a zoom-in of the three variables
at a given small time interval (red vertical lines), further confirming
the validity of Eqs.~\eqref{om_modulated} and \eqref{om_modulated_3}. We,
thus, conclude that, even under the effect of thermal fluctuations,
not only do the bending and twist degrees of freedom retain their
modulated-wave shape, but also preserve their relative phase difference.

\begin{figure}[t]
\centering\includegraphics{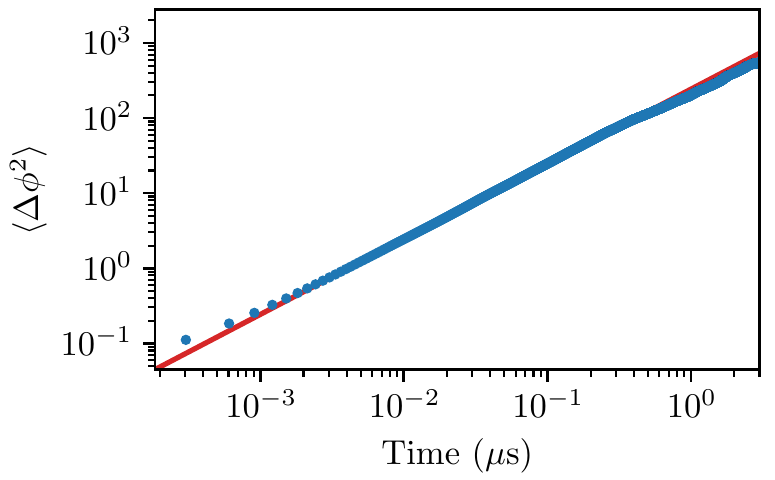}
\caption{Time evolution of the mean-squared displacement of the phase
$\phi$ (blue points), averaged over six independent oxDNA2 runs, which
reveals regular diffusion. Fitting to $\langle \Delta \phi^2\rangle =
2 D t$ (solid red line), we obtain the value $D = 121~\text{rad}^2/\mu s$
for the diffusion coefficient.}
\label{MSD}
\end{figure}

The global phase $\phi$ at the bottom of Fig.~\ref{fig:kin} was obtained
by keeping track of the orientation of the two loop ends, which yielded
the phase difference $\Delta\phi$ between successive time steps. The
initial phase was determined from the Fourier analysis of the $\Omega_2$
wave of the first frame. The mean-squared displacement of $\phi$ grows
linearly with time, as shown in Fig.~\ref{MSD}, yielding a diffusion
constant of $D = 121~\text{rad}^2/\mu s$. Note that, the simulation
time scale depends on the value of the Langevin damping parameter,
which in this case was $6.06$~ps. The origin of the observed diffusive
behavior stems from the very weak contribution of $\Delta E(\phi)$ to
Eq.~\eqref{eq:Ephi}, as discussed in Section~\ref{sec:T0}, indicating
that the phase $\phi$ moves on an almost-flat energy landscape.

\section{Conclusion} 

In this paper, we have investigated the shape and dynamics of short DNA
loops, i.e., consisting of about 100~bp.  Such loops are generated in vivo
by DNA binding proteins. We have developed a two-parameter variational
explicit solution for the minimal energy shape of the loop, which we
refer to as harmonic loop, being described by a combination of simple
trigonometric functions. This solution was found to be in excellent
agreement with numerical simulations of two different coarse-grained
models, and reproduces the exact WLC energy~\cite{yama72} up to four
significant digits.  We focused particularly on the effect of twist-bend
coupling, which is an interaction arising from the DNA grooves asymmetry
\cite{mark94}. As recently discussed in the case of DNA minicircles and
nucleosomal DNA \cite{skor18}, we found that also in DNA loops the bending
deformation induces twist waves, i.e., twist oscillations following the
periodicity of the helical pitch. Differently from the minicircle case,
in the loops discussed here twist waves have a modulated amplitude,
which is maximal at the loop apex and vanishes at the two ends. For a
loop of 100~bp the maximal degree of over- and undertwisting close to
the apex was estimated to be $6\%$ relative to the intrinsic double-helix
twist $\omega_0$. 

As an alternative approach, we could have used the numerically exact
solution by Yamakawa and Stockmayer \cite{yama72} to obtain the radius of
curvature, $R(s)$, of the loop as a function of the curvilinear coordinate
$s$. The loop shape for the model~\eqref{eq:model} is then obtained by
replacing the numerical values of $R(s)$ into Eqs.~\eqref{om_tr}. This
approach, however, would not allow for a direct estimate of derived
quantities, such as twist oscillations, curvature variation and minimal
energy.  The harmonic-loop approximation provides simple, yet accurate,
expressions for these quantities, see e.g., Eqs.~\eqref{eq:kmax},
\eqref{eq:Ephi} and \eqref{eq:kappaL2}.

Finally, we considered the loop kinetics in oxDNA simulations at room
temperature.  Interestingly, the bending and twist waves were not masked
by the presence of thermal fluctuations, and performed a correlated motion
over the whole length of the loop. This allowed us to characterize the
kinetics in terms of a single parameter $\phi(t)$, describing the absolute
phase of the waves.  This was found to follow a simple diffusive motion,
which originates from the ground-state quasi-degeneracy in $\phi$. These
findings, and particularly the simplicity of our solution, may form the
basis for more complex analytical calculations that involve strongly-bent
DNA, such as under the action of DNA-binding proteins.

\begin{acknowledgements}
We acknowledge financial support from the Research Funds Flanders (FWO 
Vlaanderen) Grant No.\ VITO-FWO 11.59.71.7N and FWO-SB 1SB4219N and
from KU Leuven grant C12/17/006.
\end{acknowledgements}

\appendix

\section{Exact variational calculus}
\label{appA}

In what follows, we briefly review the derivation of the exact solution by 
Yamakawa and Stockmayer~\cite{yama72}. Owing to the symmetry of the problem, a 
sufficient condition in order to ensure that the loop will close is to require 
that the apex of the loop has not shifted along the x-axis. Let us define 
$\vec R$ the vector connecting the end-point with the loop apex. We require that
\begin{equation} 
\vec{R} \cdot \uvec{x} = \int_0^{L/2} \uvec{e}_3 \cdot \uvec{x} \, \text ds =
\int_0^{L/2} \cos [\theta(s)] \, \text ds = 0.
\label{eq:constr}
\end{equation} 
The determination of the minimum energy under the above constraint can be 
performed by introducing a Lagrange multiplier $\mu$ as follows
\begin{equation} 
\beta E = \int_0^L \left( \frac{A \dot \theta^2}{2}
- \mu \cos \theta\right) \text ds,
\end{equation} 
where $\dot \theta \equiv \text d\theta/\text ds$. The Euler-Lagrange equation 
then becomes
\begin{equation} 
A \ddot \theta= \mu \sin \theta,
\label{eq:newt}
\end{equation} 
which has the following integral of motion
\begin{equation} 
\frac{A}{2} \dot \theta^2 + \mu \cos \theta= \Gamma.
\label{eq:tot_ene}
\end{equation} 
Interpreting $\theta$ as the coordinate of a fictitious particle with
mass $A$, and $s$ as the time variable, Eq.~\eqref{eq:newt} can be viewed
as the equation of motion for a particle in the potential $U(\theta)
= \mu\cos\theta$, describing the dynamics of a pendulum under gravity, where
$\mu$ plays the role of gravitational acceleration. This is the well-known
Kirchhoff kinetic analogy, showing that the static conformations of elastic
rods are formally equivalent to the kinetic of spinning tops.
In this analogy Eq.~\eqref{eq:tot_ene} expresses the conservation of
the mechanical energy. The boundary conditions \eqref{eq:bc} imply zero
velocity at the begin and end point, hence $\Gamma = \mu \cos \alpha$,
with $\theta(0) = \alpha$ and $\theta(L)=2\pi- \alpha$. The trajectory
$\theta(s)$ is then given by \cite{yama72}
\begin{equation}
F\left( \frac{\pi - \theta(s)}{2} , k \right) = 
F\left( \frac{\pi - \alpha}{2} , k \right) - 
\sqrt{\frac{\mu}{A}} \, \frac{s}{k},
\label{eq:FF}
\end{equation}
where $k^{-1} \equiv \cos(\alpha/2)$ and $F(\phi,k)$ is the incomplete
Elliptic integral of the first kind
\begin{equation}
F(\phi,k) = \int_0^\phi \frac{d\omega}{\sqrt{1-k^2 \sin^2 \omega}}.
\end{equation}
The values for $\alpha$ and $\mu$ are fixed by requiring that the loop
has length $L$ and that it is closed, e.g., that $\theta(s)$ satisfies
\eqref{eq:constr}.

\begin{figure}[t]
\centering\includegraphics{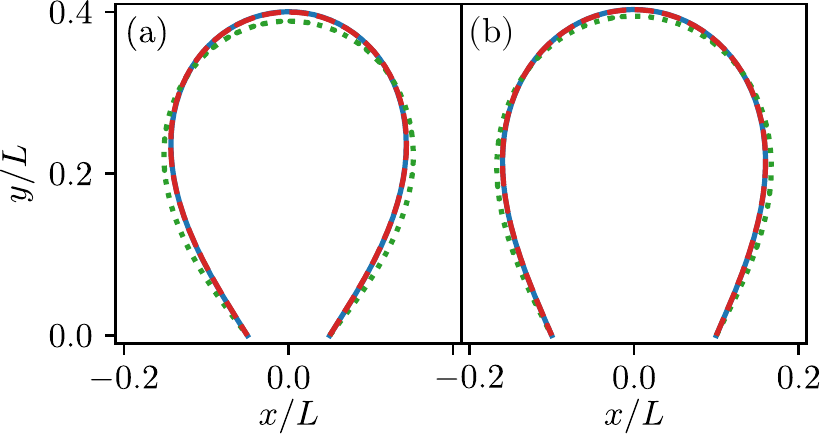}
\caption{Comparison of the shape among the one-harmonic
approximation~\eqref{eq:HL} (dotted green lines), two-harmonics
approximation~\eqref{eq:HL3} (dashed red lines) and the exact solution
(solid blue lines), with the two ends of the loop being fixed at a
distance $L/10$ (a) and $L/5$ (b).}
\label{fig:open_loop}
\end{figure}

\section{Harmonic loop ansatz}
\label{appB}

The potential energy of the pendulum has a minimum at $\theta=\pi$,
corresponding to the loop apex. As a simple approximation we expand this
potential around the minimum, which gives
\begin{equation}
U(\theta) = \mu \cos \theta \approx 
\mu \left[ -1 + \frac{(\theta - \pi)^2}{2}\right],
\label{eq:potential}
\end{equation}
corresponding to a pendulum in the small-oscillation limit. The solution
with $\theta (L/2) = \pi$ and zero velocity at the end points is then
\begin{equation}
\theta^{(1)}(s) = \pi  - c_1 \cos \left( \frac{\pi s}{L} \right).
\label{eq:theta_HL1}
\end{equation}
Note that the derivative of this solution is given by
Eq.~\eqref{eq:HL}. Plugging the above equation in \eqref{eq:constr},
and with a simple change of variables, one obtains
\begin{equation}
\int_0^{\pi/2} \cos(c_1 \cos\phi)\text d \phi \equiv 
\frac{\pi}{2} \, J_0(c_1) = 0,
\label{eq:fix_c1}
\end{equation}
where $J_0$ is the zeroth-order Bessel function of the first kind 
\footnote{The zeroth order Besssel function of the first kind has the
following integral representation
\[
J_0(x) =  \frac{1}{\pi} \int_0^{\pi} \cos (x \cos \phi) \, d\phi,
\] see \cite{abra72}, p. 360}. This constraint, thus, fixes the constant
to the smallest zero of $J_0$: $c_1 = 2.4048$.

Similarly, the third-order harmonic loop is obtained from the following
trial function
\begin{equation}
\theta^{(3)}(s) = \pi - c_1\cos\left(\frac{\pi s}{L}\right) 
- \frac{c_3}{3}\cos\left(\frac{3\pi s}{L}\right),
\label{eq:theta_HL3}
\end{equation}
with its derivative being given by Eq.~\eqref{eq:HL3}, and the
loop-closure constraint taking the form
\begin{equation}
\int_0^{\pi/2} \cos \left[ c_1 \cos\phi + 
\frac{c_3}{3} \cos(3\phi)\right] \text d \phi = 0.
\label{eq:fix_c1c3}
\end{equation}
Imposing this constraint, together with the minimization of the elastic
energy~\eqref{eq:WLC}, we obtain the values $c_1 = 2.3703$ and $c_3 =
-0.2808$ for the constants.

The above calculation can be easily generalized to open loops whose
end-points are kept at some finite distance~$d$. In that case, the form
of the harmonic solutions~\eqref{eq:theta_HL1} and \eqref{eq:theta_HL3}
remains identical, but the right-hand sides of Eqs.~\eqref{eq:fix_c1} and
\eqref{eq:fix_c1c3} are set to the nonzero value $-d/2L$ ($x$-projection
of vector pointing from the right end of the loop to the apex). As the
endpoints distance increases, the harmonic ansatz becomes a more accurate
approximation of the full solution (see Fig.~\ref{fig:open_loop}). This
is because, as $d$ increases, $\theta(s)$ varies within an interval
getting closer to $\theta =\pi$, hence the small-angle approximation
\eqref{eq:potential} becomes increasingly more accurate. This is also
reflected in the improved accuracy of the data in Table~\ref{table}
upon increasing $d$.

\section{Calculation of the energy}
\label{appC}

In this section we present some additional details over the loop-energy
calculation. For simplicity, we limit the analysis to the first-harmonic
approximation, as the third-harmonic case follows the same approach. The
energy density, obtained from Eq.~\eqref{om_modulated}, becomes (to
simplify the notation we drop the superscript in $\Omega_i^{(1)}$)
\begin{widetext}
\begin{equation} 
\frac{1}{2}\left(A_1 \Omega_1^2 + A_2 
\Omega_2^2 + C \Omega_3^2 + 
 2G  \Omega_2 \Omega_3 \right) = 
\frac{1}{2}\left( A_1 \Omega_1^2 + \widetilde A_2 \Omega_2^2 \right)
= \left( \frac{\lb \pi c_1}{L}\right)^2
\sin^2 \left(\frac{\pi s}{L}\right) 
\left[
\frac{\sin^2 (\omega_0 s + \phi)}{2A_1}
+ \frac{\cos^2 (\omega_0 s + \phi)}{2\widetilde A_2}
\right].
\label{eq:energy}        
\end{equation}
\end{widetext}
\noindent To obtain the expression in the second equality we have used
Eq.~\eqref{om_modulated} to eliminate $\Omega_3$. Interestingly, this
relation shows that the energy is identical to that of a pure bending
deformation with a reduced stiffness $\widetilde{A}_2$ instead of $A_2$.
To proceed, we use the trigonometric identities
\begin{widetext}
\begin{align}
4\sin^2 \left(\frac{\pi s}{L}\right) \sin^2 (\omega_0 s + \phi) &=
1 - \cos \frac{2\pi s}{L} -\cos \left( 2\omega_0 s + 2 \phi \right) 
+\frac{1}{2} \cos \left[2\left(\omega_0 - \frac{\pi}{L}\right)s + 2\phi \right]
+\frac{1}{2} \cos \left[2\left(\omega_0 + \frac{\pi}{L}\right)s + 2\phi \right],
\label{appC:sin} \\
4\sin^2 \left(\frac{\pi s}{L}\right) \cos^2 (\omega_0 s + \phi) &=
1 - \cos \frac{2\pi s}{L} +\cos \left( 2\omega_0 s + 2 \phi \right) 
-\frac{1}{2} \cos \left[2\left(\omega_0 - \frac{\pi}{L}\right)s + 2\phi \right]
-\frac{1}{2} \cos \left[2\left(\omega_0 + \frac{\pi}{L}\right)s + 2\phi \right],
\label{appC:cos}
\end{align}
\end{widetext}
To complete the calculation of the energy, one needs to integrate in $0 \leq s 
\leq L$. The integral of $\cos(2\pi s/L)$ vanishes; inserting 
Eqs.~\eqref{appC:sin} and \eqref{appC:cos} in Eq.~\eqref{eq:energy} and 
integrating, one finally gets
\begin{equation}
\beta E^{(1)}_\text{HL} = 
\frac{1}{2} \left( \frac{\lb \pi c_1}{L}\right)^2
\frac{L}{4} \left( \frac{1}{A_1} + \frac{1}{\widetilde{A}_2} \right)
+ \beta \Delta E(\phi).
\label{appC:Ephi}
\end{equation}
where we have separated the dominant contribution, obtained from
the integration of the constant term in the right-hand side of
Eqs.~\eqref{appC:sin} and \eqref{appC:cos}, from the part which depends on
the phase $\phi$. The former is an extensive term, giving a contribution
proportional to $L$, while the integration of the $\phi$-dependent part
gives a very small boundary contribution. Finally, using the 
definition~\eqref{eq:def_lb} of $\lb$, one sees that Eq.~\eqref{appC:Ephi} 
reduces to Eq.~\eqref{eq:Ephi} of the main text.


%
\end{document}